\documentclass[aps,pra,twocolumn,a4paper,showpacs,superscriptaddress,floatfix,10pt]{revtex4}
\usepackage[pdftex]{graphicx,color}
\usepackage{dcolumn}
\usepackage{bm}

\usepackage{amsfonts}
\usepackage{amssymb}
\usepackage{amsmath}
\usepackage{amsthm}

\usepackage{subfigure}
\usepackage{rotate}
\usepackage{dsfont}

\usepackage{pdfsync}

\let\oldmarginpar\marginpar
\renewcommand\marginpar[1]{\-\oldmarginpar[\raggedleft\tiny #1]%
{\raggedright\tiny #1}}

\def\KeyWord#1{$\backslash$\IfColor{$\!\!$\textRed{#1}\textBlack}{#1}$\!\!$}

\newcommand{\be}{\begin{equation} }
\newcommand{\ee}{\end{equation} }
\newcommand{\ba}{\begin{eqnarray} }
\newcommand{\ea}{\end{eqnarray} }

\newcommand{\mac}{\mathcal}

\newcommand{\bit}{\begin{itemize}}
\newcommand{\eit}{\end{itemize}}

\graphicspath{{../FinalFigures/}}

\begin{document}

\title{Do Rydberg chains yield Fibonacci anyons?}

\author{A. Chandran}
\email{anushyac@bu.edu}
\affiliation{Department of Physics, Boston University, MA 02215, USA}

\author{F. J. Burnell}
\affiliation{Department of Physics, University of Minnesota Twin Cities, MN 55455, USA}

\author{S. L. Sondhi}
\affiliation{Department of Physics, Princeton University, NJ 08544, USA}

\date{\today}


\date{\today}

\begin{abstract}
Recent experiments have focused attention on the properties of chains of atoms in which the atoms are either in their ground states or in highly excited Rydberg states which block similar excitations in their immediate neighbors. 
As the low energy Hilbert space of such chains is isomorphic to that of a chain of Fibonacci anyons, they have been proposed as a platform for topological quantum computation and for simulating anyon dynamics.
We show that generic local operators in the Rydberg chain correspond to non-local anyonic operators that do not preserve a topological symmetry of the physical anyons.
Consequently, we argue that Rydberg chains do not yield Fibonacci anyons and quantum computation with Rydberg atoms is not topologically protected. 
\end{abstract}

\maketitle

 \section{Introduction}

The low energy dynamics of many physical systems takes place in Hilbert spaces that \emph{do not} describe tensor products of spatially local degrees of freedom. 
Celebrated examples include the restriction to the lowest Landau level at high magnetic fields~\cite{Prange:1987aa}, and to local singlet coverings in magnetic systems~\cite{Anderson:1973aa,Rokhsar:1988kx}. 
These systems are characterized by a set of local constraints---local operators that commute with one other and with the Hamiltonian and take particular values at low energies, and are described by (generalized) gauge theories.
Gauge theories are ubiquitous in physics; in addition to describing the particle content of the universe, they underlie myriad condensed phases of matter, including superconductors, quantum Hall fluids, and spin liquids~\cite{Kogut:1979ly,Zee:2003mt,wen_book,Fradkin:2013aa}. 

Particles endowed with non-Abelian statistics emerge as the quasi-particles or topological defects of various strongly interacting quantum systems~\cite{Moore:1991hc,Read:1999tg,Kitaev:2006ab,Kitaev:2001aa,Xia:2004aa,FuKaneMajorana,OregPRL105.177002,LutchynPRL105.077001,AliceaNatPhys7,ShibaStates,BarkeshliQi,LindnerParafermion,KirillParafermion,ChengParafermion}.
They define a second class of systems with non-factorizable Hilbert spaces, in which the Hilbert space is constrained by the fusion rules that encode the outcomes of fusing pairs of particles.
Although these constrained Hilbert spaces can be isomorphic to those found in conventional gauge theories~\cite{Feiguin:2007aa,Chandran:2016ab}, the unconstrained space associated with the anyons is physically meaningless.

\begin{figure}[tb]
\centering
\includegraphics[width=\columnwidth]{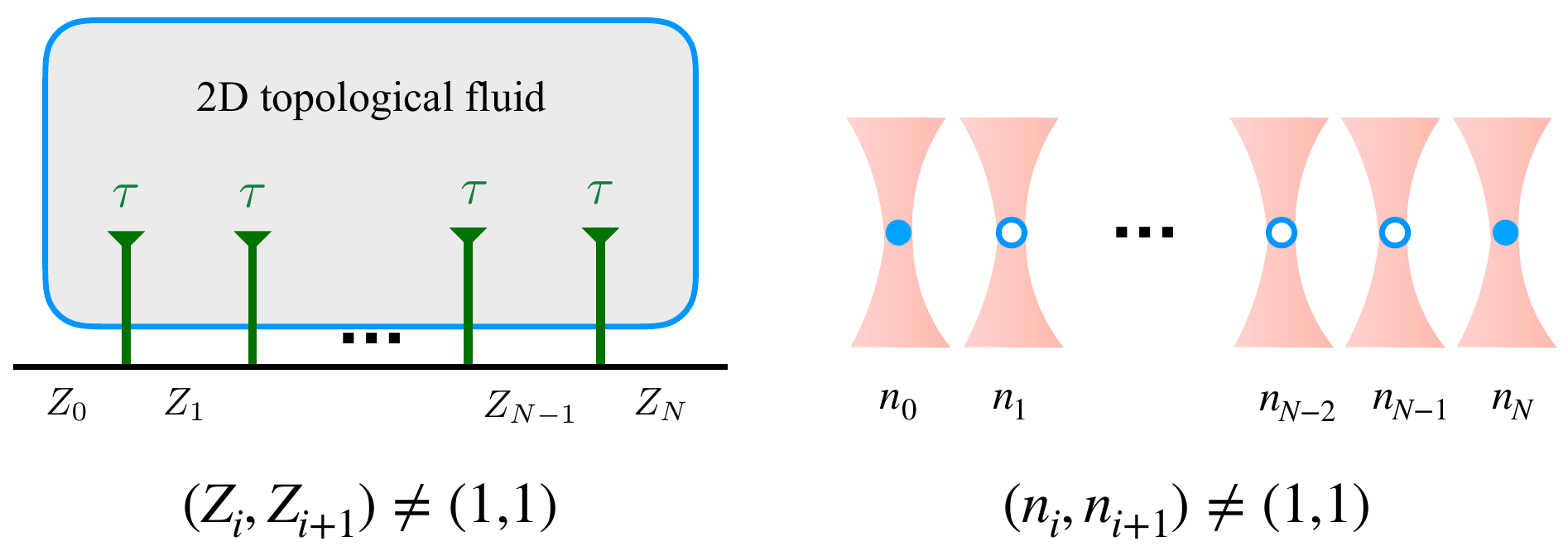}
    \caption{ Left: A chain of Fibonacci anyons in the fusion tree basis. Each vertical leg carries charge $\tau$, while the horizontal bond $i$ with label $Z_i$ represents the net fusion outcome of fusing the anyons $1, \ldots i$ with the boundary label $Z_0$. The Hilbert space is described by the labels $Z_i$ for $1\leq i \leq N$ which obey the constraint in Eq.~\eqref{Eq:FibConstraint} in each boundary condition sector $(Z_0, Z_N)$.  Right: a chain of Rydberg atoms in the blockaded regime which obey the same constraint (Eq.~\eqref{Eq:RydConst}) in each boundary condition sector $(n_0, n_N)$.  }
 \label{GCFig}
 \end{figure}

Recent experiments have created an elegant system that naively appears to fall into both classes discussed above~\cite{Schauss:2012aa,Labuhn:2016aa,Bernien:2017ab,Barredo:2018aa}. 
Specifically, the system in Ref.~\cite{Bernien:2017ab} consists of a chain of as many as $51$ neutral atoms that can support long-lived high principal quantum number ``Rydberg'' excitations. 
When the lattice constant is small enough, the low-energy thermodynamics and quantum dynamics of the Rydberg system is restricted to a locally constrained manifold by the Van der Waals interaction:
\begin{align}
\label{Eq:RydConst}
n_i n_{i+1} = 0,
\end{align}
where $n_i=0,1$ is the occupation number of a Rydberg excitation on the atom at site $i$~\footnote{This description is dual to that of a quantum dimer chain.
The quantum dimer model is in turn a strong coupling limit of a $Z_2$ lattice gauge theory~\cite{Moessner:2001tg}.}. 
While this Hilbert space clearly has a gauge theory description, it is \emph{also} isomorphic to that of a set of non-Abelian ``Fibonacci'' anyons~\cite{Read:1999tg,Feiguin:2007aa,Trebst:2008qf,Chandran:2016ab}. See Fig.~\ref{GCFig}. 
The isomorphism presents the tantalizing possibility that the Rydberg chain, like the Fibonacci chain~\cite{freedman01,Kitaev:2003vn,Stern:2008aa}, can serve as a platform for a variant of topological quantum computation and for quantum information storage~\cite{Lesanovsky:2012aa,Schulz:2013aa,Turner:2018aa,Turner:2018ab}.
Indeed, previous work~\cite{Lesanovsky:2012aa} has addressed the engineering of specific anyon Hamiltonians (such the Golden chain introduced by Ref.~\cite{Feiguin:2007aa}) with Rydberg atoms.
Should Rydberg atoms yield Fibonacci anyons, then they would further provide the first experimental realization of non-Abelian anyons, as despite many decades of experimental effort in semiconducting heterostructures~\cite{Willett:2013aa} and more recently, Indium nanowires~\cite{Mourik1003,Das:2012aa,Deng:2012aa,Churchill:2013aa,Finck:2013aa}, particles with non-Abelian statistics have proven elusive in the laboratory.  
 
In this article, we show that Rydberg chains cannot robustly simulate Fibonacci anyons for two related reasons.
First, generic local operators in the Rydberg chain are non-local in the anyon chain.
This non-locality is reminiscent of the transcription between spins and fermions in the familiar Jordan-Wigner transformation. 
The non-locality immediately implies that simulation of anyonic dynamics with Rydberg atoms has to be infinitely fine-tuned, and that the Rydberg atoms, unlike the Fibonacci anyons, do not define topologically protected q-bits.
Second, the anyonic system exhibits a topological symmetry that the Rydberg system does not.
In the boundary condition sector $n_0=n_N=0$, many operators in the Rydberg chain do not commute with this symmetry, and thus have no counterpart in the Fibonacci chain.
Altogether we conclude that the Rydberg system is properly thought of as a generalized gauge theory (see Refs.~\cite{Weimer:2010ab,Glaetzle:2014aa,Surace:2019aa}).

The outline of the article is as follows. We review the properties of Fibonacci anyons and the fusion tree basis in Section~\ref{Sec:FibAnyonReview}, before presenting the map between the Rydberg chain and the Fibonacci chain in Sec.~\ref{Sec:Mapping}. 
We then derive the topological symmetry of anyonic Hamiltonians in Sec.~\ref{Sec:TopSym} and provide examples of operators that are topologically-symmetric.
Finally, we discuss the details of trying to simulate Fibonacci anyons via Rydberg atoms and the lack of topological protection for Rydberg q-bits (Sec.~\ref{Sec:SimulFibRyd}).

\section{Fibonacci Anyons}
\label{Sec:FibAnyonReview}

Fibonacci anyons are non-Abelian particles in two dimensions~\cite{Read:1999tg,slingerland01}.
They arise as the quasi-particle excitations of certain topologically ordered fluids, e.g. the $\nu=12/5$ quantum Hall state~\cite{Xia:2004aa}.
When pinned into a one-dimensional arrangement, we obtain the Fibonacci chain discussed in Sec.~\ref{Sec:FibChain} .

We repeatedly use the process of braiding, or taking a distant anyon around a cluster of other anyons, to constrain the properties of the Fibonacci chain.
Although the braiding operation can be represented as a unitary operator in the Hilbert space of the one-dimensional Fibonacci chain, we emphasize that the physical operation can only be performed in two dimensions.
In contrast, the Rydberg chain can be measured and manipulated in one dimension.

 \subsection{Fibonacci anyons in 2 dimensions}
 \label{Sec:FibAnyonBackground}

The fundamental degrees of freedom of the Fibonacci chain are Fibonacci anyons, denoted by the symbol $\tau$.
Fibonacci anyons have two defining properties.
First, any pair of Fibonacci anyons has a net anyonic charge of be either $1$ (meaning that the two anyons can be annihilated, leaving no particles behind) or $\tau$ (meaning that if the two anyons are brought close together, they will form a single anyon of the same type).  
This total anyonic charge  --  the {\it fusion channel} of the pair -- is reminiscent of the total spin of a pair of particles; we represent the possibilities compactly through the following {\it fusion rules} \cite{rowell09}:
\begin{align}
 \label{FuseEq}
\tau \times \tau = 1 + \tau \ , \ \ \ \tau \times 1 = \tau \ , \ \ \  1 \times 1 = 1 \ \ .
\end{align}
Unlike spin, however, the total anyonic charge of any number of Fibonacci anyons necessarily takes on one of only two values, $1$ or $\tau$.  

The second defining property of Fibonacci anyons is their anyonic statistics.  Specifically, \emph{braiding} one Fibonacci anyon around another in two dimensions leads to a net phase that depends on the total fusion channel of the two anyons.
This braiding operation is non-Abelian in the sense that it is a matrix-valued operation in the 2-dimensional Hilbert space of the pair of $\tau$ anyons.
In Sec.~\ref{Sec:ProjMake}, we describe how this matrix-valued operation can be used to realize the operator that projects onto the vacuum fusion channel of a cluster of $\tau$-anyons.

We note that a system of non-Abelian anyons in two dimensions is equivalent to a system of bosons with a `statistical interaction' that is chosen such that the phases due to braiding are accrued under adiabatic exchange (for a detailed discussion, see Ref.~\cite{BondersonThesis, Trebst:2008qf}). 


\subsection{Fibonacci chain and the fusion tree basis}
\label{Sec:FibChain}
The Fibonacci chain is composed of a line of ${N}$ {$\tau$-anyons} with labels $1, \ldots N$.
In Fig.~\ref{GCFig}, the vertical legs represent the anyons; the anyonic charge of each leg is $\tau$.  
Each state in the Hilbert space of the Fibonacci chain can be specified by a set of labels $Z_i$ for $i = 1, \ldots N-1$ on each horizontal bond.  The $Z_i$ basis is called the fusion tree basis, because
the $Z_i$ labels satisfy the fusion rules in Eq.~\eqref{FuseEq}: 
\begin{align}
Z_{i+1} = \tau \times Z_i \label{Eq:Zbasis}
\end{align}
Since the trivial anyonic charge combined with the $\tau$ anyon always gives a $\tau$ anyon, the Hilbert space consists of all assignments of $Z_i$ obeying the constraint that no two consecutive bonds take the value $1$:
\begin{align}
(Z_i, Z_{i+1}) \neq (1,1) \quad \textrm{for }0 \leq i \leq N-1. \label{Eq:FibConstraint}
\end{align}
We define $\hat{Z}_i$ to be the operator that measures the label $Z_i$ on bond $i$:
\begin{align}
\hat{Z}_i \equiv \left\{
\begin{array}{cc}
1, & Z_i = 1 \\
-1, & Z_i = \tau
\end{array}
\right.
\end{align}

For open chains, the values of $Z_1$ and $Z_{N-1}$ are further constrained by the boundary conditions -- i.e. by the values of $Z_0$ and $Z_{N}$ in Fig. \ref{GCFig}.   
For example, if $Z_0 = 1$, then we must have $Z_1 = Z_0 \times \tau = \tau$, whereas if $Z_0 = \tau$, we have $Z_1 = \tau \times \tau \in \{ 1, \tau \}$.  
Physically, $Z_0$ is the net fusion channel of other anyons to the left of the Fibonacci chain, while $Z_{N}$ is the net channel of the anyons to the right. 
The four possible boundary condition sectors, $(Z_0, Z_N) = (1,1), (1,\tau), (\tau,1)$ and $(\tau,\tau)$, have different Hilbert space dimensions:
\begin{align}
 \label{Eq:HSize}
n_N = F_{N-1+R(Z_0) + R(Z_N)}\stackrel{N \rightarrow \infty}{ \approx} \frac{\phi^{N-1+R(Z_0) + R(Z_N)}}{\sqrt{5}}
\end{align}
where $F_k$ is the $k^{th}$ Fibonacci number, $R(1) = 0$, $R(\tau) =1$, and
\begin{align}
\phi = \frac{1}{2} \left( 1 + \sqrt{5} \right )
\end{align}
is the {\it Golden mean}. 

In deriving Eq.~(\ref{Eq:HSize}) --and indeed throughout this work --  we assume that the net fusion channel of the anyons in the chain (green vertical legs in Fig.~\ref{GCFig}) with the two boundary labels $Z_0$ and $Z_{N}$ is one.  
This constrains the total topological charge of the Fibonacci chain to be:
\begin{align}
\textrm{Net topological charge} = Z_0 \times Z_N
\end{align}
In the $(1,1), (1,\tau), (\tau,1)$ boundary condition sectors, the net topological charge is unique and respectively given by $1$, $\tau$ and $\tau$.
If $Z_0 = Z_N = \tau$ however, the net charge can either be $1$ or $\tau$. 

In what follows, we treat the anyons in the Fibonacci chain as point particles because they are well-separated relative to the correlation length of the topological fluid that they are embedded in.

\subsection{Projectors onto fusion channels}
\label{Sec:ProjMake}

We now discuss how braiding processes in the two-dimensional topological fluid hosting the Fibonacci chain determine the nature of the operators that can act on the chain.   
We first describe how braiding a probe $\tau$ particle realizes the operator $P^1_N$ that projects a cluster of $N$ Fibonacci anyons onto the vacuum fusion channel.  We then discuss projectors involving a subset of the anyons in the chain.

\subsubsection{Realizing projectors through braiding}
\label{Sec:BraidingProj}

 \begin{figure}[tb]
\centering
\includegraphics[width=0.8\columnwidth]{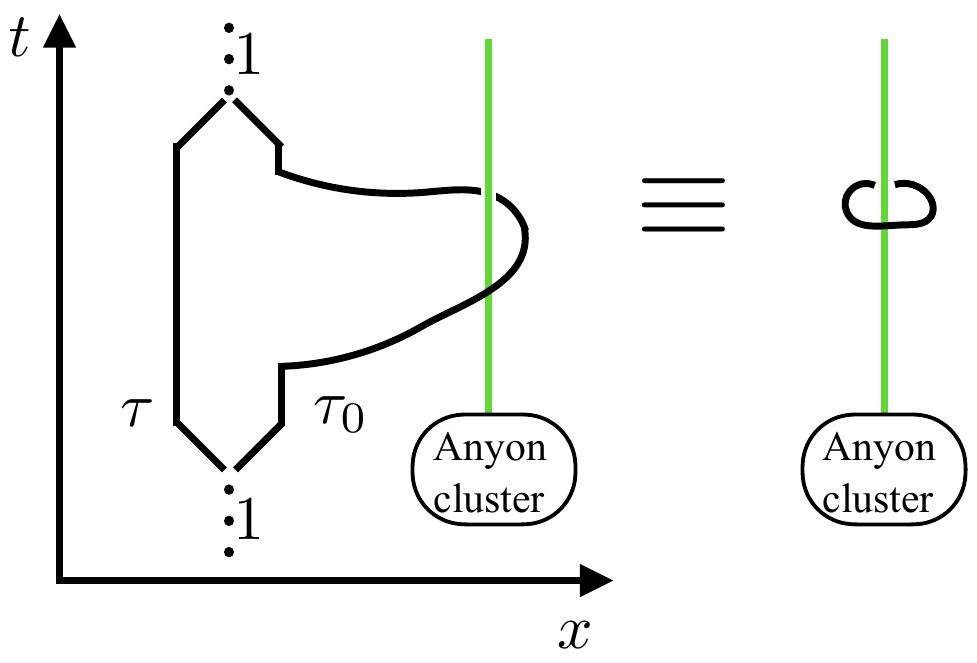}
    \caption{Left: Space-time depiction of braiding process that realizes ${P}^{1}_{N}$. A pair of Fibonacci anyons is created from the vacuum extremely far from the anyonic cluster of interest.  One member of the pair ($\tau_0$) is taken around the cluster, remaining at all times extremely far away from any anyons in the cluster.  The pair is then brought back together and annihilated.  Right: By deforming the world lines, the process is represented as a closed loop encircling the anyon cluster.
 }
 \label{BraidFig}
 \end{figure}

Consider a pair of probe $\tau$ anyons created from the vacuum far away from the cluster of anyons forming the Fibonacci chain.  
Braid one of the probe anyons $\tau_0$ around the cluster keeping $\tau_0$ extremely far from the cluster at all times, such that only the long-ranged statistical interactions contribute to the accumulated phase.  
Then return the system to its original state by annihilating the probe particles into the vacuum. A space-time depiction of the entire process is shown in Fig.~\ref{BraidFig}.  Note that  
by deforming the world lines, the process can be represented as a closed loop encircling the anyon cluster (Fig. \ref{BraidFig}); this representation will be particularly useful in Sec.~\ref{Sec:TopSym}. 

Using the basic fusion and braiding data of Fibonacci anyons in 2D~\cite{BondersonThesis}, it is easy to show that the above process is described by the operator \cite{BondersonThesis}
\be \label{Eq:Otest}
O_{\mathrm{test}} = |1 \rangle \langle 1 | - \phi^{-2} |\tau\rangle \langle \tau| 
\ee
where $|i\rangle$ represents the state of the cluster with net fusion channel $i = 1, \tau$.  
 We emphasize that the outcome of the process depends crucially on the fact that we project the two probe anyons onto the vacuum both before and after the braiding experiment.  Different choices of the initial and final fusion channel for this pair will result in different coefficients in front of the $|\tau \rangle \langle \tau |$ term in Eq.~\eqref{Eq:Otest}.  
 For a discussion of these more general statistical interactions, see Ref. \cite{BondersonThesis}.

Intuitively, Eq. (\ref{Eq:Otest}) tells us that if the cluster has a net topological charge of $1$, it is indistinguishable from the vacuum at long distances and the statistical interaction between $\tau_0$ and the cluster is zero.
As the direct anyon-anyon interactions decay exponentially in the distance between the anyons in a gapped topological fluid, their contribution to the accumulated phase of the joint wavefunction of the cluster and the probe anyons in the braid operation also vanishes as the separation between $\tau_0$ and the cluster becomes large.
Thus, braiding $\tau_0$ around the cluster is equivalent to braiding it around the vacuum, and the probability that the probe particles annihilate into the vacuum after the braid operation is one. 

If the cluster has a net topological charge of $\tau$ on the other hand, then the long-ranged statistical interaction with $\tau_0$ 
can change the net fusion channel of the pair of probe anyons.
In this case, the probability that the probe anyons fuse to the vacuum after the braid is less than one.

Define ${P}^{m}_{N} = |m \rangle \langle m |$ to be the operator that projects the $N$-anyon cluster onto a state with net charge $m$ for $m=1,\tau$. From Eq.~(\ref{Eq:Otest}) we see that:
\begin{align}
O_{\mathrm{test}} &= {P}^{1}_{N} - \phi^{-2}{P}^{\tau}_{N} \\
& = (1+ \phi^{-2}) {P}^{1}_{N} - \phi^{-2} \mathds{1}
 \end{align}
The braid experiment with probe $\tau$ particles can thus be used to realize the operator ${P}^{1}_{N}$ that projects an $N$ anyon cluster onto the vacuum fusion channel.  Notice that $P_N^1$ is the net topological charge of the entire Fibonacci chain, and is a c-number fixed by the boundary conditions in all but the $Z_0 = Z_N = \tau$ boundary condition sector.


 \begin{figure}[tb]
\centering
\includegraphics[width=\columnwidth]{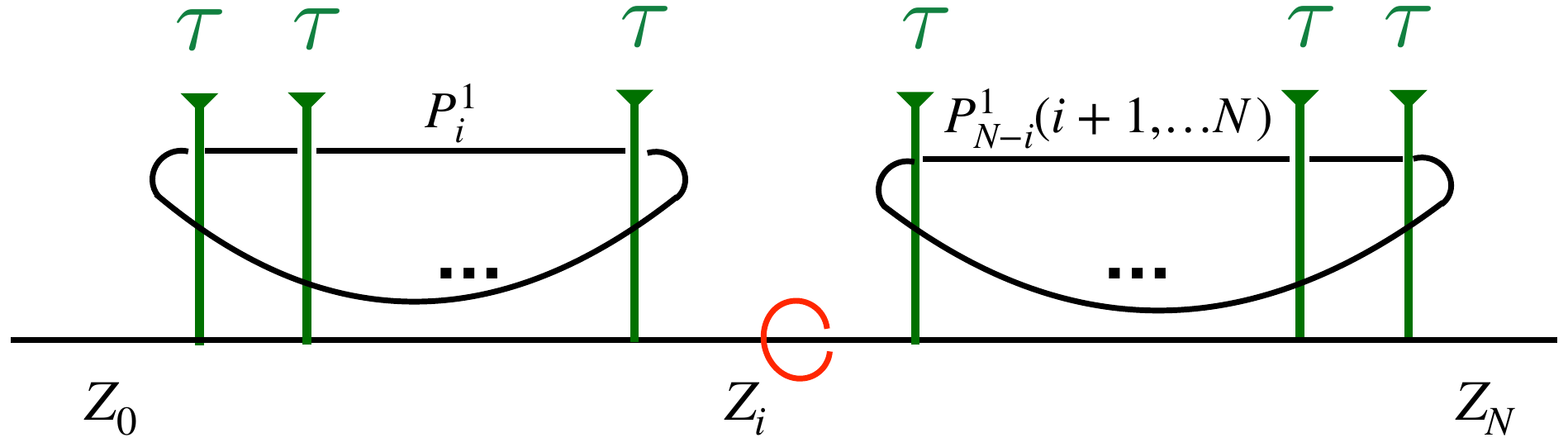}
    \caption{Pictorial depiction of various operators of the Fibonacci chain. The operator $P_i^1$ projects the fusion outcome of the first $i$ anyons in the Fibonacci chain to the vacuum channel, while $P_{N-i}^1(i+1, \ldots N)$ projects the outcome of the anyons $i+1, \ldots N$. The value of $Z_i$ can be measured through a braid experiment around the boundary leg $Z_0$ and the anyons $1, \ldots i$,  or  through a braid around the boundary leg $Z_N$ and the anyons $i+1, \ldots N$. By deforming the world-lines of either braid experiment, we obtain the loop encircling the bond $i$.
 }
 \label{ProjFig}
 \end{figure}

\subsubsection{Projectors onto fusion channels of subsets of anyons}

Braiding the probe $\tau_0$ anyon around a subset of anyons in the Fibonacci chain defines the projection operator that projects the subset into a specific fusion channel \footnote{The definition through braiding requires the subset of anyons to be far away from the remaining anyons in the chain.}.
Let $\mac{P}^\alpha_{n}(i_1, ... i_n)$  denote the operator that projects the cluster of $n$ anyons $i_1, ... i_n$ onto the fusion channel $\alpha = 1, \tau$.  
For example, $\mac{P}^1_2(i,j)$ projects anyons $i$ and $j$ into the vacuum fusion channel. 
When the anyon site labels are suppressed, the projector acts on all the anyons \emph{in the chain} to the left of the index in the subscript.
Thus, the symbol $P_i^1$ is short-hand for the following operator,
\begin{align}
P^1_i = \mac{P}^1_{i}(1, \ldots i),
\end{align}
which projects the fusion outcome of the first $i$ anyons in the Fibonacci chain to the vacuum. See Fig.~\ref{ProjFig}.

\subsection{Relationship to $\hat{Z}_i$ operators}
The projectors $P^1_i$ are closely related to the $\hat{Z}_i$ operators in certain boundary condition sectors. First, note that $P^1_i$ does not act on the boundary label $Z_0$. 
In contrast, we have
\begin{align}
Z_1 = \tau_1 \times Z_0 \ , \ \ Z_2 = \tau_2 \times Z_1 \ , \ \ ...   
\end{align} 
so that $Z_i$ can be interpreted as the net fusion channel of anyons $1, \ldots i$ with the boundary label $Z_0$:
\begin{align}
Z_i = \tau_i \times ( \tau_{i-1} \times ( \ldots (\tau_1 \times Z_0 ))) \label{Eq:Zidef}
\end{align}

If $Z_0 = 1$, then $Z_i$ is simply the outcome of fusing anyon $i$ with anyons $1, ... i-1$.  
Consequently, in the $(1,1)$ and $(1,\tau)$ boundary condition sectors:
\begin{align}
\textrm{Topological charge of anyons }& 1, \ldots i  = Z_i \\
\Rightarrow P_i^1 &= \frac{1+\hat{Z}_i}{2} \label{Eq:ZandP11}
\end{align}

In the $(\tau, 1)$ boundary condition sector, a similar relation holds because of the identity $Z_0 \times \tau_1 \times ... \times \tau_i = \tau_{i+1} \times \tau_{i+2} \times ... \times Z_{N}$.  
As
\begin{align}
Z_i = \tau_{i+1} \times ( \tau_{i+2} \times ( \ldots \times (\tau_{N} \times   Z_{N}  ) )), \label{Eq:Zidef2}
\end{align}
and $Z_{N} = 1$, we have:
\begin{align}
\textrm{Topological charge of anyons } i+1, \ldots N = Z_i
\label{Eq:ZandP1tau}
\end{align}
and $P_{N-i}^1(i+1, \ldots N) = (1+\hat{Z}_i)/2$.   

When $Z_0 = Z_N = \tau$, $Z_i$ is not equivalent to the net fusion outcome of anyons in the Fibonacci chain alone (see Eqs.~\eqref{Eq:Zidef},~\eqref{Eq:Zidef2}). 
In this case, the operator $\hat{Z}_{i}$ acts on the boundary labels and is not completely determined by $P_i^1$ or $P_{N-i}^1(i+1, \ldots N)$ (although it commutes with both projectors).

\section{The map between Rydberg-blockaded and Fibonacci chains}  \label{Sec:Mapping}
As summarized in Fig.~\ref{GCFig}, the states in the fusion tree basis are in one-to-one correspondence with the occupation number states of the nearest neighbor Rydberg-blockaded chain in each boundary condition sector if we identify:
\begin{align}
|Z_i = 1 \rangle \Leftrightarrow |n_i = 1\rangle \nonumber\\
|Z_i = \tau \rangle \Leftrightarrow |n_i = 0\rangle \label{Eq:Identify}
\end{align}
The local constraint in Eq.~\eqref{Eq:FibConstraint} then corresponds to perfect Rydberg blockade Eq.~\eqref{Eq:RydConst}.

The mapping does not preserve operator locality. Specifically, \emph{local operators in the Rydberg chain} generically map to {\it non-local} operators in the Fibonacci chain. This is reminiscent of the one-dimensional Jordan-Wigner transformation in which local spin operators that anti-commute with parity are mapped to string operators in the Majorana basis.

For example, consider the operator $\hat{n}_i$, which measures the occupation number of atom $i$ in the Rydberg chain.  
From Eq.~\eqref{Eq:Identify}, we find: 
\begin{align}
\frac{1}{2} ( 1 + \sigma^z_i ) \equiv n_i \Leftrightarrow\frac{1}{2} (1+\hat{Z}_i)   \label{Eq:zitoni}
\end{align}
As discussed in Sec. \ref{Sec:FibChain}, the operator $\hat{Z}_i$ encodes the net fusion channel of anyons  $1, .. i$ with the boundary label $Z_0$ (or equivalently, the fusion channel of anyons $i+1, ... N$ with the boundary label $Z_N$) and thus requires $i$ or $(N-i)$-body measurements.

Similarly, the operator that flips the Rydberg occupation number on site $i$, $\sigma_i^x$, maps to a sum of local and non-local operators in the fusion tree basis:
\begin{align}
\sigma_i^x	\Leftrightarrow   &\phi^{3/2}\mac{P}^{1}_{2}(i, i+1)  + \frac{\sqrt{\phi}(1-\phi)}{4}  (\hat{Z}_{i-1} + \hat{Z}_{i+1}) \nonumber \\
 &- \frac{\phi^{5/2}}{4} \hat{Z}_{i-1} \hat{Z}_{i+1} - \frac{(1-\phi)}{2\sqrt{\phi}}\hat{Z}_{i}  \label{Eq:SigXintermsofP}
 \end{align}
where we have suppressed an additive constant. Although the first term in the RHS is a local projector in the anyon chain, the remaining terms measure the net fusion channel of all the anyons to the left of $i-1$ or $i+1$ with the boundary labels, and thus make the entire RHS non-local.

Interestingly, \emph{local operators in the Fibonacci chain} map to \emph{local} operators in the Rydberg chain. 
For example, the projector onto the vacuum fusion channel of anyons $i$ and $i+1$ maps to:  
\begin{align}
 \mac{P}^{1}_{2}(i, i+1)	\Leftrightarrow   \frac{ \sigma_i^x}{\phi^{3/2}} - \phi^{-1} (n_{i-1} + n_{i+1}-1) \nonumber \\
 + \phi n_{i-1} n_{i+1} + \frac{(1-\phi)}{\phi^{2}}n_{i} 
 \end{align}
using Eqs.~\eqref{Eq:zitoni} and ~\eqref{Eq:SigXintermsofP}.
More generally, consider an interaction term of range $m$ involving anyons $k$, $k+m-1$ and any number of anyons in-between.
The interaction term cannot modify the fusion outcome of fusing any collection of anyons with indices in $\{1,\ldots, k-1, k+m, \ldots N\}$ as it does not act on any of these anyons.
More stringently, the interaction term cannot \emph{measure} fusion outcomes associated with these anyons except for their net fusion channels. 
Let $Y_{\ell}$ denote the net fusion channel of the group of anyons to the left $\{1,\ldots, k-1\}$ and $Y_{r}$ that of the group to the right $\{k+m,\ldots, N\}$; our projector is diagonal in $Y_{l,r}$.
As $Z_{k-1}$ ($Z_{k+m-1}$) is the result of fusing $Y_\ell$ ($Y_r$) with the boundary leg $Z_0$ ($Z_N$), we conclude that the interaction term is diagonal in the $Z_{k-1}, Z_{k+m-1}$ basis and in general, depends on the values of $Z_{k-1}, Z_{k+m-1}$.
Note that the interaction term acts as the identity on the bonds $1, \ldots k-2, k+m, \ldots N$ because it cannot measure the internal structure of the left and right groups.
Thus, the interaction term at most connects states that differ in their $Z_k \ldots Z_{k+m-2}$ labels for any choice of $Z_{k-1}, Z_{k+m-1}$, and maps to a local $(m+1)$-body operator in the Rydberg chain.

\section{Topological symmetry of anyonic Hamiltonians} \label{Sec:TopSym}
In this section, we argue that the Fibonacci chain's total topological charge must be conserved under its own dynamics, while the Rydberg chain has no analogous conservation law.  
We call the principle underlying this conservation ``topological symmetry", because it follows from the braiding and fusion rules of the two-dimensional fluid that hosts the Fibonacci anyons. 
We also discuss the relationship to the topological symmetry of the Fibonacci chain with periodic boundary conditions~\cite{Feiguin:2007aa}.

Suppose that the Fibonacci chain's total topological charge is ill-defined.
Then, a generic eigenstate can be decomposed as:
\begin{align}
|E\rangle = {P}^1_N |E\rangle + P^\tau_N |E\rangle
\end{align}
Next, consider performing the operation described in Sec.~\ref{Sec:ProjMake}.
That is, consider creating a pair of $\tau$ particles from the vacuum infinitely far away from the chain, bringing one of these around the system at a rate that is sufficiently slow that it does not create any excitations along its path, and re-annihilating the pair into a vacuum state.
From Eq.~\eqref{Eq:Otest}: 
\begin{align}
O_{\mathrm{test}} |E \rangle = -\frac{1}{ \phi^2} P^\tau_N |E\rangle + P^1_N |E\rangle,
\end{align}

Suppose the state $|E \rangle$ is non-degenerate.
Since $O_{\mathrm{test}} |E \rangle \neq |E \rangle$, the process of braiding a test particle infinitely far away from the chain changes its energy. 
This is clearly impossible for any physical system.
Thus non-degenerate eigenstates $|E\rangle$ must have a definite fusion outcome.   

Suppose the eigenspace at energy $E$ is degenerate.
Then, $O_{\mathrm{test}} |E \rangle$ is a linearly independent eigenstate with the same energy as $|E \rangle$.
That is, we can always diagonalize the fusion outcome within the degenerate eigenspace. 
Thus, in either case, we conclude that like total spin, the total fusion outcome of our system is conserved under the dynamics: 
\begin{align}
[H, {P}^1_N]=0 \label{Eq:TopSymm}
\end{align}
where $H$ is the Hamiltonian of the system.
This is the topological symmetry.

A few comments are in order. 
First, the Hamiltonian of a Fibonacci chain is composed of operators that conserve the total topological charge, much as the Hamiltonian of a spin chain with spin-rotational symmetry is composed of terms that conserve the total spin.  
Unlike in the spin system, however, the topological symmetry does not imply spectral degeneracy.  In a $SU(2)$-symmetric spin system, a fixed total spin $S_{\text{tot}} >0$ requires that the system as a whole transform in a $2S_{\text{tot}} +1$-dimensional representation of $SU(2)$.   
Fibonacci anyons have no analogue of these representations; an isolated $\tau$ particle has no internal quantum numbers apart from its total anyon charge.   
Correspondingly, Fibonacci anyons permit no raising/lowering operators that commute with $H$.

Second, Ref.~\cite{Feiguin:2007aa} identified a ``topological symmetry" associated with a particular operator that commutes with $H$ on a chain with periodic boundary conditions.  Though the operator in question  (or, more specifically, its square) is related to the braiding process described above, it is physically distinct and represents a process that is specific to the periodic Fibonacci chain.  In particular, the topological symmetry with periodic boundary conditions implies a 2-fold degeneracy in the spectrum.  Nonetheless both operators capture the conserved quantity that results from the topological order of the two-dimensional fluid that the anyons are embedded in.

\subsection{Number of linearly independent topologically symmetric operators}
\label{Sec:Nlin}
The topological symmetry restricts the operators that enter into the Fibonacci chain's Hamiltonian in certain boundary condition sectors.
We begin here by counting the total number of operators that obey topological symmetry. 

In the $(1,1)$, $(1,\tau)$ and $(\tau,1)$ boundary condition sectors, the topological symmetry places no constraints on the anyonic system's Hamiltonian because $P^1_N$ is a c-number. 
In the $(\tau,\tau)$ sector however, Eq.~\eqref{Eq:TopSymm} imposes that the Hamiltonian is block-diagonal on the two possible fusion outcomes. 
To see that this reduces the number of possible operators entering the Hamiltonian, note first that any topologically symmetric operator $O$ can be expressed as:
\begin{align}
O &=   P_N^1 O  {P}_N^1 +   P_N^\tau O  P_N^\tau \\
&= \hat{O}_1 + \hat{O}_\tau
\end{align}
The number of linearly independent operators $\hat{O}_1$ is the number of linearly independent operators in the Fibonacci chain with $(1,1)$ boundary conditions because the fusion tree basis with $Z_{0} = Z_N = 1$ is a basis for the Hilbert space with $P_N^1=1$.
Similarly, the number of linearly independent operators $\hat{O}_\tau$ is the number of linearly independent operators in the $(1,\tau)$ or $(\tau,1)$ sectors.
Thus, from Eq.~\eqref{Eq:HSize}, the total number of linearly independent topologically symmetric operators acting on an anyonic chain with $N$ anyons is:
\begin{align}
n_\mathrm{op} = F_{N-1}^2 + F_{N}^2 \label{Eq:nop}
\end{align}
We observe that $n_{\mathrm{op}}$ is less than the number $F_{N+1}^2$ of linearly independent operators in the $(\tau,\tau)$ boundary condition sector. Thus, not all Hermitian matrices in the $(\tau, \tau)$ sector represent valid anyonic Hamiltonians. 

\subsection{Examples of topologically symmetric operators}
We now investigate which operators commute with the topological symmetry.  We show that any projector $\mac{P}^1_m(i_1, ... i_m)$ that projects onto the vacuum fusion channel for any subset of $m$ anyons commutes with the total topological charge, while operators such as $\hat{Z}_i$ that describe the net fusion channel of anyons in the chain with the boundary labels do not.  As the topological symmetry acts non-trivially only in the $(\tau,\tau)$ boundary condition sector, here we restrict our attention to this case. 

\begin{figure}[htp]
\centering
\includegraphics[width=\columnwidth]{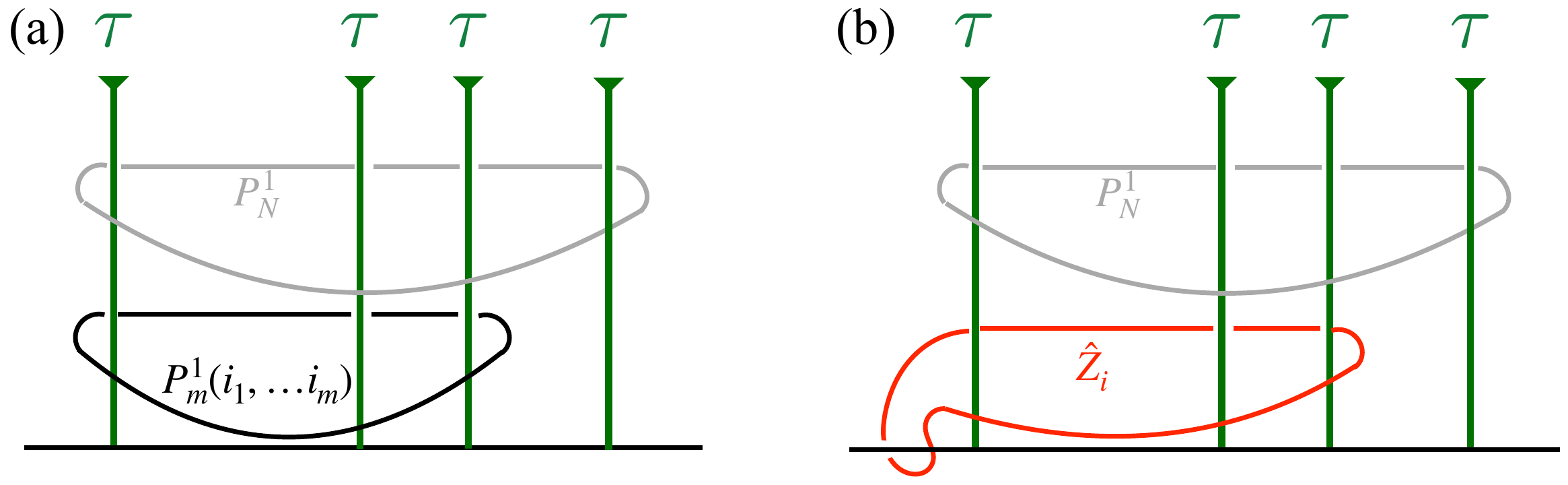}
    \caption{ Illustration of loops representing different braiding operations on the Fibonacci chain.  As explained in detail in Ref.~\cite{Nayak2008}, loops that can slide past each other freely (such as those associated with the projectors $P^1_N$ and $P^1_m(i_1, ... i_m)$ shown in panel (a)) commute, while loops that cannot slide freely past each other (such as those associated with $P^1_N$ and $\hat{Z}_i$, see panel (b)) do not. }
 \label{Fig:Loops}
 \end{figure}

Recall that the projector $\mathcal{P}^1_{m}(i_1, \ldots i_m)$ can be carried out by first separating anyons $i_1, \ldots i_m$ from the other anyons in the chain, and then braiding a probe anyon around this subset of anyons (see Sec.~\ref{Sec:BraidingProj}).   
An important feature of the braiding process is that it commutes with $P_1^N$. 
A diagrammatic ``proof" follows from the representation of $\mathcal{P}^1_{m}(i_1, \ldots i_m)$ as a loop around the world-lines of the anyons with labels $i_1, \ldots i_m$ (see Fig.~\ref{Fig:Loops}), on noting that loops in the space-time representation that are able to freely pass through one another indicate commuting projectors. (Conversely, loops that cannot slide past one another represent measurements that do not commute).
For further details on the mathematics underlying the diagrammatic calculus, see Ref.~\cite{Nayak2008}. 
As the loop around the world lines of all $N$ anyons can be passed through any loop encircling a subset of the anyons:
\begin{align}
[P_N^1, \mathcal{P}^1_{m}(i_1, \ldots i_m)]=0
\end{align}
Thus, projectors onto a given fusion channel of a subset of anyons are topologically symmetric operators. The analogue of this statement in the $SU(2)$-symmetric case is that the total angular momentum of any subset of the spins respects the global $SU(2)$ symmetry. 

The next question is whether these projectors constitute a complete basis for all operators compatible with topological symmetry.    
Here simple counting arguments do not suffice: a super exponential in $N$ number of topologically symmetric operators can be constructed by taking tensor products of the different projection operators $\mathcal{P}^1_{m}(i_1, \ldots i_m)$.  
These operators must be linearly dependent, as their number exceeds the total number of linearly independent topologically symmetric operators $n_{\mathrm{op}} \sim \phi^{2N}$ (see Eq.~\eqref{Eq:nop}).  
Nevertheless, we expect that these operators span the topologically symmetric operator space, as they completely specify the information about the state of the system accessible by measurements on the anyons in the chain.  
Indeed, in the $(1,1), (\tau,1)$, and $(1, \tau)$ boundary condition sectors, we can use the mapping between Rydberg operators $\sigma^x_i$ and $\sigma^z_i$ and non-local projectors (Eqs.~\eqref{Eq:ZandP11},~\eqref{Eq:ZandP1tau},~\eqref{Eq:zitoni},~\eqref{Eq:SigXintermsofP}) to construct a basis of topologically symmetric operators in terms of projectors.   
Thus, any topologically symmetric anyonic Hamiltonian can be expressed as a sum of products of projection operators $\mathcal{P}^1_{m}(i_1, \ldots i_m)$.

With $(\tau,\tau)$ boundary conditions, the operators corresponding to the local Rydberg operations  $\sigma^z_i$ and $\sigma^x_i$ are not topologically symmetric.  The space-time diagram in Fig.~\ref{Fig:Loops}(b) provides the diagrammatic proof. From Eq.~\eqref{Eq:zitoni}, $\sigma^z_i$ maps to the projector associated with a braiding process that encircles the boundary label $Z_0$, as well as the anyons in the chain with labels $1, \ldots i$.  
The resulting loop cannot freely slide past a loop encircling all anyons in the chain; hence 
\begin{align}
[\sigma_i^z, P^1_N] \neq 0.
\end{align}
Similarly, using the expression for $\sigma^x_i$ in Eq.~\eqref{Eq:SigXintermsofP} in terms of projectors that encircle the boundary leg with label $0$, it is easily shown that $[\sigma_i^x, P_1^N]\neq0$. We note that finely tuned combinations of operators in the Rydberg chain can be topologically symmetric if they are an algebraic combination of projectors onto a subset of the anyons in the chain.

\section{Simulating Fibonacci anyons with Rydberg chains}
\label{Sec:SimulFibRyd}


\subsection{Hamiltonians with $(1,1), (1,\tau)$, and $(\tau,1)$ boundary conditions}

In Sec.~\ref{Sec:TopSym}, we showed that all operators in the constrained Hilbert space conserve the total topological charge of the chain in the $(1,1)$, $(1,\tau)$, and $(\tau,1)$ boundary condition sectors.  However, in Sec.~\ref{Sec:Mapping}, we pointed out important differences in locality between operators in the Rydberg and Fibonacci chains.  For example, in the $(1,1)$ boundary condition sector, the local Rydberg operator $n_i \equiv \frac{1}{2} ( 1+ \sigma^z_i )$ is represented in the Fibonacci model by $P_i^1$, which projects all anyons to the left of bond $i$ into the vacuum fusion channel.  The operator $\sigma^x_i$ similarly maps to sums of non-local projectors (see Eq. (\ref{Eq:SigXintermsofP})).  Superficially, these differences in locality are reminiscent of those arising in the mapping between the 1D Ising chain and the 1D Majorana chain~\cite{Kitaev:2001aa}.  However, there is an important distinction: in the latter case, Ising symmetric operators are local in both representations, such that aspects of the dynamics of one model can be usefully studied with the other.  In the case at hand, there appears to be no symmetry (or other reasonable restriction) on operators in the Rydberg chain such that they produce only local operators in the anyon chain.  As a consequence, any realistic attempt to create a Fibonacci-type Hamiltonian with Rydberg atoms~\cite{Lesanovsky:2012aa} must be the result of fine-tuning, such that only the specific linear combinations of products of $\sigma^x$ and $\sigma^z$ that are local in the anyon model appear. Deviation from this fine-tuning results in an effective Hamiltonian for the Fibonacci chain that is non-local.

\subsection{Hamiltonians with $(\tau,\tau)$ boundary conditions}

With $(\tau,\tau)$ boundary conditions, the net topological charge of the anyon chain can be $1$ or $\tau$, and topological symmetry imposes non-trivial constraints on the Hamiltonians.  Since the Hilbert space has dimension $F_{N+1}$ for an $N$-anyon chain, the total number of operators in the Rydberg Hilbert space is $F_{N+1}^2$; however as discussed in Sec.~\ref{Sec:Nlin}, only $F_{N}^2 + F_{N-1}^2$ of these are compatible with conservation of topological charge.  These include the projectors $P^i_m(i_1..., i_m)$.

Thus, in addition to the important differences in locality between Fibonacci and Rydberg Hamiltonians, the spectrum of a Fibonacci chain is block diagonal in the total topological charge, while the spectrum of generic Rydberg Hamiltonian is not. This has striking consequences for the energy spectrum of a Fibonacci chain:
\begin{enumerate}
\item The energy spectrum with $(\tau, \tau)$ boundary conditions is a direct sum of the energy spectra with $(1,1)$ and $(1,\tau)$ boundary conditions.
\item The energy spectrum with $(1, \tau)$ boundary conditions is identical to that with $(\tau, 1 )$ boundary conditions.
\end{enumerate}
These properties of the Fibonacci spectrum could be exploited to test whether a given Rydberg chain is successfully emulating a chain of non-abelian anyons.

\subsection{Consequences for topological quantum computing}

One of the most important motivations for constructing quantum simulators that emulate the Fibonacci chain is to exploit the potential of Fibonacci anyons for universal topologically protected quantum computation~\cite{Kitaev:2003vn,Nayak2008,Bonsteel05,hormozi07,hormozi09}.   We thus discuss some specific architectures for Fibonacci q-bits, their analogues in the Rydberg picture, and the fate of topological protection in the face of random noise in the Rydberg Hamiltonian.

\begin{figure}[htp]
\centering
\includegraphics[width=0.4\columnwidth]{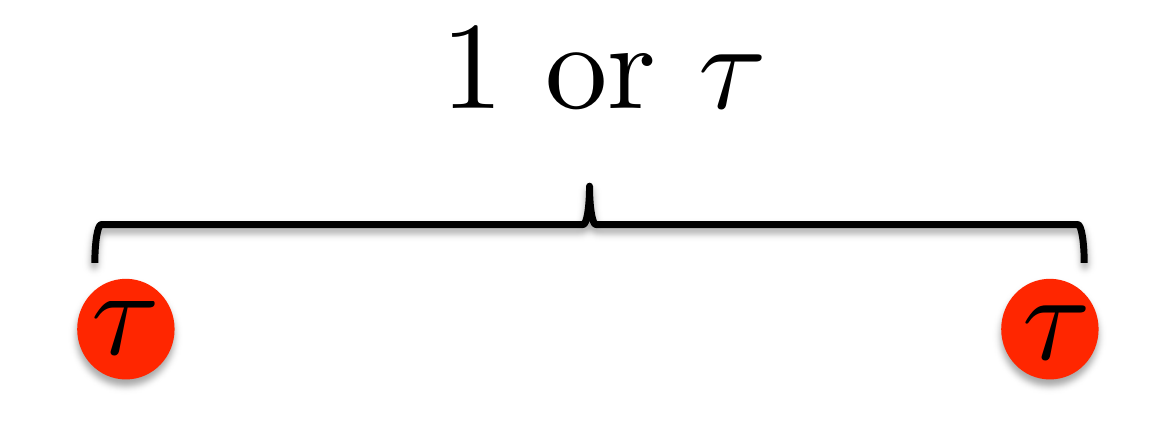}
\caption{\label{QbitFig} An example of a q-bit composed of one segment of Fibonacci chain, with two well-separated Fibonacci anyons.  Information is encoded in the net topological charge of the 2-anyon system, which can take on values of $1$ or $\tau$.}
\end{figure}

Any Fibonacci chain segment with two or more anyons defines a q-bit by associating the states of total topological charge $1$ and $\tau$ with the two $z$-states of the q-bit. (Evidently, each chain segment must be in the $(\tau,\tau)$ boundary condition sector to be able to represent both possible fusion outcomes).  The topological charge of each segment is conserved by any Hamiltonian involving only the anyons on that particular segment; hence the information is topologically protected in as much as different segments can be isolated from each other.  Additionally, if the anyons within the segment are weakly interacting, then the energy splitting between the two fusion channel outcomes of each segment can be very small.  For example, if the segment contains two well-separated Fibonacci anyons, the splitting between the $1$ and $\tau$ fusion channels is expected to be exponentially small in the separation.  A q-bit of this type is shown in Fig.~\ref{QbitFig}.  

One of the appealing features of non-abelian anyons for applications in quantum computing is that, at least in theory, it is possible to manipulate the state of the q-bit through non-local (braiding) processes, whereas all local operators preserve the state of the q-bit.  For the 2-anyon q-bit, the braiding process brings a third anyon between the pair involved in the q-bit (which we assume to be well-separated relative to any relevant correlation lengths); such a process has some amplitude of flipping the state of the q-bit from $1$ to $\tau$ (and vice versa).  Fibonacci anyons are a particularly exciting platform for this type of quantum computing since braiding operations can be used to realize a universal set of quantum gates -- in contrast to platforms involving Majorana zero modes, which cannot~\cite{MajoranaTQC} .

However, there are several obstacles to using Rydberg atoms to simulate Fibonacci q-bits.  First, there is no topological symmetry leading to a conserved fusion outcome for each chain segment of Rydberg atoms.  In particular, the on-site Rydberg operators $\sigma^x_i$ and $\sigma^z_i$ for $i$ within the segment fail to conserve the segment's topological charge.  Thus, the q-bit's state is affected by random noise in these on-site terms. Second, since the braiding operation is a feature of the anyons in two dimensions, the topologically protected manipulation of quantum information is also not easily achievable with local Rydberg Hamiltonians. Specifically, the braiding process between two anyons separated by a distance $m$ maps to a $(m+1)$-body fine-tuned unitary gate on the Rydberg atoms; any noise on the unitary gate would decohere all the q-bits defined by the $(m+1)$ Rydberg atoms. 

\section{Concluding Remarks}

In closing, we have described the connection between two different physical problems with
isomorphic Hilbert spaces---those of the Rydberg chain characterized by local constraints and of Fibonacci anyons characterized by anyonic fusion rules. 
The mapping is different from the more familiar Jordan-Wigner mapping between spins and fermions in one dimension in several respects.  
Specifically, we find that operators that are local in the Fibonacci chain are also local in the Rydberg chain, while local Rydberg operators generically map to {\it non-local} anyonic operators.  
In addition, the Fibonacci chain has a topological symmetry arising from its embedding in a two-dimensional topologically ordered fluid, which has no natural analogue in the Rydberg model.
Finally, we showed that the non-locality of the mapping between the spaces implies that q-bits in the Rydberg chain do not have the desired topological protection of quantum information.  
Irrespective of the application to anyonic simulators, we expect the fascinating interplay of theory~\cite{Fendley:2004aa,Sela:2011aa,Vasseur:2015ab,Turner:2018aa,Lin:2019aa,Moudgalya:2018aa,Ostmann:2019aa,Feldmeier:2019aa,Verresen:2019aa,Khemani:2019aa,Iadecola:2019aa} and experiment~\cite{Schauss:2012aa,Labuhn:2016aa,Bernien:2017ab,Barredo:2018aa} to uncover many new features of constrained systems using the Rydberg atomic simulator. 

\section{Acknowledgements}
We are grateful to M. Zaletel for several discussions on the connections between the Rydberg-Fibonacci map and the Jordan-Wigner map.
We are also grateful to C.R. Laumann for valuable discussions on anyons.
This work was supported by NSF DMR-1752759 (A.C.), NSF-DMR-1352271 (F.J.B) and the US Department of Energy grant No. DE-SC0016244 (S.L.S.).
A.C. acknowledges support from the Sloan Foundation through the Sloan Research Fellowship.

\bibliography{paper-master}

\end{document}